\documentclass[a4paper,11pt]{article}
\usepackage{jheppub} % for details on the use of the package, please see the JINST-author-manual
\usepackage{slashed}
\usepackage{amsmath}
\usepackage{amssymb}
\usepackage{amsthm}
\usepackage{hyperref}
\usepackage{indentfirst}
\usepackage{psfrag}
\usepackage{graphicx}
\usepackage{cleveref}
\Crefname{figure}{Fig.}{Figs.}% {<type>}{<singular>}{<plural>}
\usepackage{subcaption}
\usepackage[utf8]{inputenc}
\hypersetup{colorlinks=true, linkcolor=blue, urlcolor=blue, citecolor=blue, linktocpage=true}
\usepackage{enumerate}
\usepackage[dvipsnames]{xcolor}
\usepackage{color}

\makeatletter
\def\@fpheader{\relax}
\makeatother

\usepackage{xifthen}% Provides \isempty test
\newcommand*\diff{\mathrm{d}} % Straight differential
\newcommand*\ldiff[2][]{ \ifthenelse{\isempty{#1}}{ \diff #2}{\diff^#1#2} \,} % Differential with measure; the mandatory argument is the name of the measure, the option one is the dimension

\def\e{{\rm e}}
\def\d{\partial}

\newcommand{\td}[1]{\tilde{#1}}

\newcommand\vf{\varphi}
\newcommand{\bvf}{\bar{\varphi}}
\newcommand{\bt}{\bar{\tau}}
\newcommand{\br}{\bar{r}}
\newcommand{\bV}{\bar{V}}
\newcommand{\tr}{\td{r}}
\newcommand{\tdt}{\td{\tau}}

\renewcommand{\l}{\lambda}
\newcommand{\eq}[1]{\begin{align}#1\end{align}}

\usepackage{lineno}
\title{\huge{Holographic phase transitions via thermally-assisted tunneling}}

\author[a]{Tony Gherghetta,}

\author[a]{Arpon Paul,}

\author[b]{and Andrey Shkerin}
\emailAdd{tgher@umn.edu}
\emailAdd{paul1228@umn.edu}
\emailAdd{ashkerin@perimeterinstitute.ca}
\affiliation[a]{School of Physics and Astronomy, University of Minnesota, 116 Church St SE, Minneapolis, Minnesota 55455, USA}
\affiliation[b]{Perimeter Institute for Theoretical Physics, Waterloo, Ontario N2L 2Y5, Canada }

\abstract{We construct the thermal bounce solution in holographic models that describes first-order phase transitions between the deconfined and confined phases in strongly-coupled gauge theories. This new, periodic Euclidean solution represents transitions that occur via thermally-assisted tunneling and interpolates between the $O(4)$-symmetric vacuum bubble at zero temperature and the high temperature $O(3)$-symmetric critical bubble associated with classical thermal fluctuations. The exact thermal bounce solution can be used to obtain the bounce action at low temperatures which allows for a more accurate determination of vacuum decay rates, significantly improving previous estimates in holographic models.
In particular, provided the phase transition is sufficiently supercooled, new predictions are obtained for the gravitational wave signal strength for critical temperatures ranging from the TeV scale up to $10^{12}$ GeV, some of which are within reach of future gravitational wave detectors.}

\begin{document}
\begin{flushright}
{\small UMN–TH–4419/25}\\
\end{flushright}
\maketitle
\flushbottom

\section{Introduction}
\label{sec:Intro}

Cosmological phase transitions play an important role during the evolution of the early universe \cite{Mazumdar:2018dfl}.
Besides the QCD and electroweak phase transition, which have been well studied, there could be first-order phase transitions associated with supersymmetry breaking, composite Higgs dynamics, axion and neutrino physics or the GUT scale. A first-order phase transition is particularly interesting because the transition proceeds via bubble nucleation where the subsequent bubble collisions are sufficient to generate gravitational waves. The remnant of this energetic process today is a stochastic gravitational wave (GW) background \cite{Caprini:2015zlo,Caprini:2018mtu,Caprini:2019egz,Athron:2023xlk} which could be detected in future observations.

In finite-temperature Yang--Mills theories the transition can occur from a deconfining phase at high temperatures to a confining phase at low temperatures. Since these transitions occur at strong coupling they can be studied using holography where they have a dual gravitational description via the AdS/CFT correspondence. In particular, models based on a compact warped extra dimension~\cite{Randall:1999ee} provide a simple, well-motivated framework to study these holographic phase transitions
\cite{Creminelli:2001th,Randall:2006py,Nardini:2007me,Konstandin:2010cd,Bunk:2017fic,Dillon:2017ctw,Megias:2018sxv,Baratella:2018pxi, Agashe:2019lhy,Fujikura:2019oyi,Megias:2020vek,Bigazzi:2020phm,Agashe:2020lfz,Mishra:2023kiu,Mishra:2024ehr,Agrawal:2025wvf} (see also \cite{DelleRose:2019pgi,VonHarling:2019rgb,Fujikura:2025iam}).
In these models parameters controlling the breaking of the conformal symmetry in the IR are small, which is necessary for explaining the large hierarchy between the Planck and the weak scales. 
This leads to a large potential barrier separating the phases and, as a result,
the phase transition may be delayed until the universe experiences a large amount of supercooling. 
When the phase transition eventually occurs, this delay causes 
a much larger energy release, and therefore supercooled transitions feature an enhanced GW signal, which makes them appealing from the phenomenological point of view.

A crucial characteristic of first-order phase transitions is the time of the most probable nucleation of the bubble of true vacuum. This moment is found by equating the temperature-dependent decay rate of the metastable phase with the Hubble rate of expansion.
In turn, the decay rate evaluated using the saddle-point approximation is
\begin{equation}\label{Gamma}
\Gamma(T)=A(T)\e^{-B(T)} \;,
\end{equation}
where the temperature dependence of both the exponential suppression factor 
$B$ (obtained from the Euclidean action) and the prefactor $A$ has been emphasized. At small $T$, the decay proceeds via quantum tunneling amplified by thermal fluctuations, while at sufficiently high temperatures, $T>T_q$, the decay occurs due to classical thermal fluctuations, where $T_q$ is the transition temperature between the quantum and classical regimes \cite{Weinberg:2012pjx}. 
In the standard picture of vacuum decay based on the imaginary-time formalism \cite{Coleman:1977py,Callan:1977pt}, quantum tunneling is described by the time-dependent periodic solution of the equations of motion (or thermal bounce), while classical transitions are described by the time-independent critical bubble \cite{Linde:1980tt,Linde:1981zj}. 
Assuming that the metastable state exists at all temperatures $T<T_q$, then when the temperature varies between zero and $T_q$, the thermal bounce interpolates between the $O(4)$-symmetric vacuum bubble (obtained at $T=0$) and the $O(3)$-symmetric critical bubble.

In this paper, we explore holographic phase transitions in the regime where thermally-assisted tunneling dominates over classical decays.
By considering the holographic model studied in Refs.~\cite{Agashe:2019lhy,Agashe:2020lfz,Mishra:2023kiu,Mishra:2024ehr}, we show that the quantum decay channel is preferred for certain parameters of the model, thus allowing new parameter regions to be explored. 
Working in the 4D effective theory for the radion field, we construct the thermal bounce numerically and study its behavior at different temperatures. At low $T$ the bounce exhibits the $O(4)$-symmetry and resembles the vacuum bubble.\footnote{In the holographic model that we study, the thermal bounce cannot be smoothly connected with the zero-temperature bubble, since the limit $T\to 0$ is outside the range of validity of the 4D theory.} As the temperature grows, the shape of the bounce is deformed and its time-dependence diminishes. 
Interestingly, we find that the holographic thermal bounce exists in some range of temperatures above $T_q$ where its Euclidean action exceeds that of the critical bubble.\footnote{In 4D theories in flat space, this behavior is known to occur in cases when the bounce potential is close to the thin-wall limit \cite{Garriga:1994ut,Ferrera:1995gs}.}

Our work fills an important gap in the analysis of holographic first-order phase transitions. It has been customary to use $O(4)$-symmetric configurations as an approximation to the thermal bounce at low $T$ and to determine $T_q$ as the temperature at which the Euclidean action $B_0$ of this configuration equals the critical bubble action $B_c$.\footnote{That this is an approximation can be understood by inspecting the boundary conditions: the $O(4)$-symmetric configuration does not satisfy the condition of periodicity in the imaginary time required by thermal equilibrium.} 
This simple approximation is based on the assumption that the $O(4)$-symmetric configuration resembles the true thermal bounce at all temperatures $T<T_q$. However, even the existence of such a bounce interpolating between the $O(4)$-symmetric and $O(3)$-symmetric solutions is not obvious given the fact that the boundary conditions for the Euclidean solutions in the holographic setup differ from those normally imposed in flat space.
In fact, the standard picture of vacuum decay is modified in many important cases and in curved spacetime. For example, there are no time-dependent, periodic bounces around many types of black holes in thermal equilibrium \cite{Briaud:2022few}. Nevertheless, the fact that we have been able to obtain the thermal bounce solution for a particular class of holographic models allows for a more complete study of holographic phase transitions.

In particular, finding the thermal bounce is important for making accurate predictions for observables associated with the phase transition, such as the stochastic GW background. Under generic assumptions of model parameters, including a cubic term in the Goldberger--Wise potential, we compute the gravitational wave signal that arises from supercooled phase transitions occurring at temperatures $T\lesssim T_q$, which can potentially be detected in future gravitational wave observatories such as LISA \cite{LISA:2017pwj,Baker:2019nia}, BBO \cite{Crowder:2005nr,Corbin:2005ny,Harry:2006fi,Yagi:2011wg} and DECIGO \cite{Seto:2001qf,Kawamura:2006up,Isoyama:2018rjb}.
These transitions are achieved for critical temperatures ranging from the TeV scale up to $10^{12}$~GeV, which is particularly interesting due to the relevance for holographic transitions associated with the electroweak hierarchy, dynamical supersymmetry breaking~\cite{Buyukdag:2018ose,Buyukdag:2018cka} and composite axion models~\cite{Cox:2021lii}.

The paper is organized as follows. 
In section~\ref{sec:setup} we review the holographic setup to study the phase transition and the 4D effective theory for the radion field. Working within this theory, in section~\ref{sec:sol} we construct the relevant thermal bounce at $T<T_q$ and compare its Euclidean action with that of the critical bubble.
In section~\ref{sec:transit} we calculate the nucleation temperature $T_n$ marking the onset of the phase transition for different model parameters and explore the regime $T_n<T_q$ where quantum tunneling dominates the transition. In section~\ref{sec:pheno} we discuss the GW signal generated by collisions of bubbles produced in the quantum regime and our conclusion is given in section~\ref{sec:concl}. 
Several appendices contain the analysis of linear perturbations around the critical bubble, the numerical procedure to find the bounce and a more detailed study of the thermal bounce near the transition temperature.

\section{Review of the holographic setup}
\label{sec:setup}

We will work in the holographic picture of the deconfined-confined phase transition in a finite-temperature gauge theory, when both phases are at strong coupling. 
The description includes a 5D anti-de Sitter--Schwarzschild (AdS-S) spacetime that is truncated with a UV and IR brane. At high temperatures the theory is in the deconfined phase, where the IR brane is hidden behind the AdS-S (or black brane) event horizon and the conformal symmetry is unbroken. 
This geometry is described by the following 5D metric with coordinates $(t,\vec{x}, \rho)$~\cite{Agashe:2020lfz}
\begin{equation}
\label{BBMetric}
    \diff s^2 = -\left( \rho^2 - \frac{\rho_h^4}{\rho^2}\right)\diff t^2 + \frac{\diff \rho^2}{\rho^2-\frac{\rho_h^4}{\rho^2}} + \rho^2\sum_{i=1}^3\diff x_i^2 \,,
\end{equation}
where the extra-dimensional coordinate $\rho$ satisfies $\rho_h<\rho<\rho_{\rm UV}$ and  $\rho_h(\rho_{\rm UV})$ are the locations of the event horizon (UV brane) (assuming $\rho_h\ll \rho_{\rm UV}$).
Note that we set the AdS curvature length, $\ell_{\rm AdS}=1$, as in \eqref{BBMetric}, unless explicitly needed. 
The surface $\rho=\rho_h$ corresponds to the event horizon which is related to the Hawking temperature associated with the AdS-S solution by
\begin{equation}
    T=\frac{\rho_h}{\pi}\,.
    \label{eq:HawkingT}
\end{equation}
Since the IR boundary is hidden behind the event horizon, the Hawking temperature \eqref{eq:HawkingT} corresponds to the temperature of the deconfined phase in the dual 4D gauge theory.

As the temperature drops, the theory transitions to the confined phase described by a slice of AdS 
between the two branes and the conformal symmetry is spontaneously broken. The corresponding 5D geometry is
\begin{equation}\label{RSmetric}
    \diff s^2 = -\rho^2\diff t^2 + \frac{\diff\rho^2}{\rho^2} + \rho^2\sum_{i=1}^3\diff x_i^2\,,
\end{equation}
where $\rho_{\rm IR} < \rho < \rho_{\rm UV}$ and $\rho_{\rm IR}$ is the location of the IR brane.
The two phases coexist at the critical temperature $T_c$. The free energy barrier separating the phases leads to a first-order phase transition when the bubbles of the confined phase appear in the metastable, deconfined phase, at the nucleation temperature $T_n<T_c$.

To fix the location of the IR brane in the confined phase, one introduces a Goldberger--Wise (real) scalar field $\chi$ with bulk and brane potentials~\cite{Goldberger:1999uk}.
The full action of the 5D theory is
\begin{equation}
S_5=\int \diff^5x\sqrt{-g}\left( -2M_5^3R-\Lambda_5 \right)-\sum_{i={\rm UV, IR}}\int\diff^4x\,\sqrt{-g_i}\,{\cal T}_i +S_\chi\;,
\label{eq:S5}
\end{equation}
where $M_5$ is the 5D Planck scale, $\Lambda_5$ is the bulk cosmological constant, ${\cal T}_i$ are brane tensions satisfying $-\Lambda_5={\cal T}_{\rm UV}=-{\cal T}_{\rm IR}=24M_5^3$ 
and $g_i$ is the induced metric\footnote{Note that in \eqref{eq:S5} we have not explicitly written the Gibbons--Hawking--York boundary term which is also needed to have a well-defined action with boundary terms~\cite{Agashe:2020lfz}.}. The action of the Goldberger--Wise scalar field is
\begin{equation}
S_\chi = \int\diff^5x\sqrt{-g}\left( -\frac{1}{2}(\d\chi)^2-U(\chi) \right) - \sum_{i={\rm UV, IR}}\int\diff^4x\sqrt{-g_i}U_i(\chi) \,.
\end{equation}
By the AdS/CFT correspondence, the 5D gravitational theory is dual to a strongly-coupled 4D gauge theory, where the 5D Planck scale $M_5$ is related to the number of colors $N_c$ in the 4D gauge theory via the relation
\begin{equation}
    (M_5\ell_{\rm AdS})^3 \approx \frac{N_c^2}{16\pi^2}.
    \label{eq:M5Nc}
\end{equation}
However, the AdS curvature scale $1/\ell_{\rm AdS}$ cannot exceed the cutoff scale of the 5D gravitational description which is estimated to be $(24\pi^3)^{1/3} M_5$~\cite{Giudice:2003tu}. Thus, imposing this condition and using \eqref{eq:M5Nc} we obtain $N_c\gg 1$ or equivalently $1/\ell_{\rm AdS} \ll 40 M_P$~\cite{ParticleDataGroup:2024cfk}, where $M_P^2\simeq \frac{1}{2} M_5^3 \ell_{\rm AdS}$ with $M_P=2.4\times 10^{18}$~GeV. To satisfy this constraint on the number of colors we will assume $N_c\geq 3$.

The bulk potential $U(\chi)$ includes a symmetry-breaking mass term and, possibly, nonlinear terms. The latter terms enhance the conformal symmetry breaking in the IR. In particular, following Ref.~\cite{Mishra:2023kiu}, we consider a cubic self-interaction of $\chi$, so that the bulk potential for the Goldberger--Wise field is 
\begin{equation}\label{GWpot}
U(\chi)=2\epsilon_2\chi^2+\frac{4}{3}\epsilon_3\chi^3 \;,
\end{equation}
where $\epsilon_2,\epsilon_3<0$. The signs of $\epsilon_{2,3}$ are chosen so that the corresponding deformation in the dual gauge theory is a relevant operator that becomes larger at IR scales.
{We also assume $0.01 \lesssim|\epsilon_2|,|\epsilon_3|\lesssim 0.1$ ensuring that the Goldberger-Wise profile $\chi\left(\rho\right)$ varies slowly across the bulk and remains sensitive to both branes, which is necessary for generating a radion potential capable of stabilizing a large hierarchy: $\rho_{\rm IR}\ll\rho_{\rm UV}$~\cite{Goldberger:1999uk,Mishra:2023kiu}. }
The boundary potentials $U_i(\chi)$ are chosen as follows
\begin{equation}
\begin{aligned}
& U_{\rm UV}(\chi)=\beta_{\rm UV}(\chi-v_{\rm UV})^2 \;, ~~ \beta_{\rm UV}\to\infty \;, \\
& U_{\rm IR}(\chi)=2\alpha_{\rm IR}\chi \,,
\end{aligned}
\label{eq:boundaryU}
\end{equation}
which determine the boundary conditions for the field $\chi$.
Using \eqref{eq:boundaryU} they are given by $\chi(\rho_{\rm UV})=v_{\rm UV}$ and $\rho_{\rm IR}\chi'(\rho_{\rm IR})=-\alpha_{\rm IR}$.
Note that in the deconfined phase, 
the $\chi$ boundary condition is imposed on the event horizon instead of the IR brane.

The potential (\ref{GWpot}) and the boundary conditions determine the bulk profile $\chi(\rho)$.
The profile depends on the potential parameters $\epsilon_2, \epsilon_3, v_{\rm UV}, \alpha_{\rm IR}$ as well as the parameters of the 5D solution $\rho_{\rm IR}$ and $T$.\footnote{One can rescale $\rho$ to set the position of the UV brane to be at $\rho_{\rm UV}=1$.}
Knowing the profile, one can calculate the 4D effective action for the radion field $\vf$ (with $\langle\vf\rangle\equiv\rho_{\rm IR}$),
\begin{equation}\label{eq:Seff}
S=\frac{3N_c^2}{2\pi^2}\int\diff^4x\sqrt{-g}\left( -\frac{1}{2}(\d\vf)^2-V(\vf) \right) \,,
\end{equation}
where we have used the relation \eqref{eq:M5Nc} to show that the effective action is proportional to $N_c^2$.
The radion potential has the form
\begin{equation}
\label{EffPot}
V(\vf)=\kappa^4\vf^4+\frac{2\pi^2 }{3N_c^2}\vf^4\left(a_2\,\frac{\l\vf^{\epsilon_2}}{1-\l\vf^{\epsilon_2}}-a_3\log\left(1-\l\vf^{\epsilon_2} \right)\right)\;,
\end{equation}
where $\kappa\lesssim 1$ is a parameter that encodes a small back-reaction effect of the Goldberger-Wise field on the IR brane tension and $\lambda,a_2,a_3$ are related to the 5D parameters as follows \cite{Mishra:2023kiu}:
\begin{equation}
\l=\frac{1}{1+\epsilon_2/(v_{\rm UV}\epsilon_3)} \;, ~~
a_2=-\frac{1}{32}\epsilon_2\alpha_{\rm IR}^2-\frac{\epsilon_2}{\epsilon_3}\alpha_{\rm IR}+2\alpha_{\rm IR} \;, ~~
a_3 = \frac{\epsilon_2}{2\epsilon_3} \alpha_{\rm IR}\;.
\end{equation}
The second term in the potential \eqref{EffPot} breaks the conformal symmetry and mimics the effects of strong dynamics.

\begin{figure}[t]
    \begin{minipage}[t]{0.48\textwidth}
         \includegraphics[width=\linewidth]{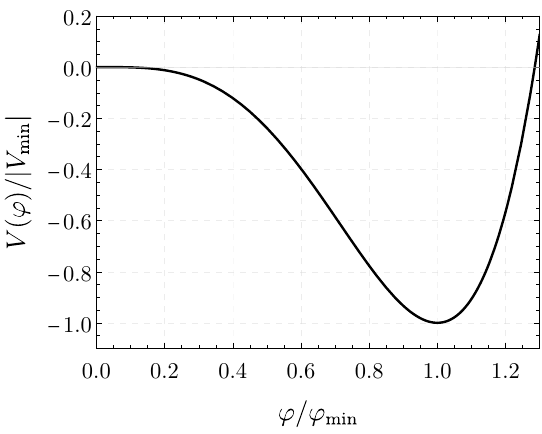}
    \end{minipage}
  \hfill
    \begin{minipage}[t]{0.48\textwidth}
         \includegraphics[width=\linewidth]{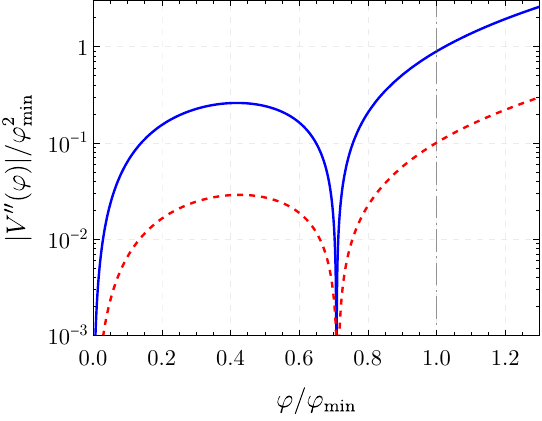}
    \end{minipage}
   \caption{\textit{Left:} The radion potential (\ref{eq:rad_pot_approx}) rescaled to a form which is independent of the model parameters. 
   \textit{Right:} The magnitude of the curvature of the potential scaled by $\vf_{\rm min}^2$, assuming 
   $m_\vf^2=0.1$ (red dashed line) and $m_\vf^2=0.9$ (blue solid line).
   The radion mass is determined from the value of the curvature at $\vf = \vf_{\rm min}$, shown as the dot-dashed vertical line.
   The kink at $\vf/\vf_{\rm min}\approx 0.7$ corresponds to an inflection point, where the curvature changes sign from negative to positive as $\vf$ increases.
}
    \label{fig:rad_pot}
\end{figure}

The potential \eqref{EffPot} develops a singularity at $\vf_s=\l^{-1/\epsilon_2}$. This singularity is an artifact of the analytical approximation to the solution $\chi(\rho)$ \cite{Mishra:2023kiu} and limits the validity of the 4D theory to the region $\vf\gg\vf_s$. 
In this region, the potential has a minimum at $\vf=\vf_{\rm min}$. This minimum corresponds to the true vacuum of the theory at temperatures $T<T_c$.
At $\vf_s\ll\vf\lesssim\vf_{\rm min}$ and $|\epsilon_2|\ll 1$, the potential can be conveniently approximated as
\begin{equation}\label{eq:rad_pot_approx}
    V(\vf)\approx \frac{m_\vf^2}{16}\vf^4\left( -1 + 4 \log\left( \frac{\vf}{\vf_{\rm min}} \right) \right) \;,
\end{equation}
where $m_{\vf}^2 = V''\left(\vf_{\rm min}\right)/\vf_{\rm min}^2$ is 
the radion mass squared at the minimum in units of $\vf_{\rm min}^2$.
Rescaling $V(\vf)$ and $\vf$ by $|V_{\rm min}|\equiv |V(\vf_{\rm min})|= m_\vf^2\vf_{\rm min}^4/16$ and $\vf_{\rm min}$, respectively, causes the shape of the potential to be 
independent of the parameter values.
The rescaled potential is shown in the left panel of figure~\ref{fig:rad_pot}. Note the potential is not valid at $\vf/\vf_{\rm min}\lesssim  \vf_s/\vf_{\rm min} \ll 1$ (not discernible on the plot).

For convenience, we will work with the parameters $|V_{\rm min}|$ and $m_{\vf}^2$ in what follows, since $|V_{\rm min}|$ is directly related to the critical temperature, as will be explained in section~\ref{sec:sol}, while $m_{\vf}^2$ controls the curvature of the potential and so determines the quantum-to-classical transition temperature $T_q$. In particular, we will consider two benchmark values of the radion mass squared: $m_\vf^2 = 0.1$ and $0.9$. These are shown in the right panel of figure~\ref{fig:rad_pot} and correspond to the parameter values $(\kappa^4,\epsilon_2,\epsilon_3) = (0.25,-0.1,0)$ and $(0.1,-0.04,-0.12)$
of the potential (\ref{EffPot}), respectively. 
It is worth noting that the radion mass squared is given by $m_\vf^2=-4\kappa^4\epsilon_2$ for $\epsilon_3 = 0$ and is further enhanced when the cubic coupling $\epsilon_3$ is nonzero\footnote{It can be numerically checked that $m_\vf^2$ also depends on $\alpha_{\rm IR}$ and $v_{\rm UV}$ for non-zero $\epsilon_3$.}.
The remaining parameters, $\alpha_{\rm IR}$, $v_{\rm UV}$, will be adjusted to the desired value of $|V_{\rm min}|$ keeping $m_\vf^2$ fixed to the benchmark values.
Finally, we assume $N_c=3,6$ as the benchmark numbers of colors entering the overall factor in the action (\ref{eq:Seff}).\footnote{For all numerical computations,
$N_c$ is first conveniently set to unity and then the obtained action is scaled by the appropriate value of  $N_c^2$.}

At temperatures $T<T_c$ the confined phase is the true vacuum and it is described within the 4D theory (\ref{eq:Seff}). 
On the other hand, the metastable deconfined phase is not described by the 4D theory, since $\vf$ is not defined in this phase. Thus, in discussing the phase transition, it is crucial to understand the limits of validity of the 4D theory and to ensure that the dominant contribution to the bounce action is obtained from where the 4D description is applicable. Besides, one should understand the boundary condition on the 4D bounce imposed by matching with the deconfined phase (as will be shown in section~\ref{sec:sol}).
One of the validity conditions of the 4D theory is that temperature corrections to the potential (\ref{EffPot}) from excitations of the Goldberger--Wise and Kaluza--Klein fields are small. The corrections scale linearly with the temperature, hence we require \cite{Agashe:2020lfz,Mishra:2023kiu}
\begin{equation}\label{CondField}
   \vf_s \ll T \ll \vf \;,
\end{equation}
where the lower bound comes from the validity of the potential (\ref{EffPot}).
Also, to ensure that the back-reaction of the Goldberger--Wise field profile on the 5D geometry is negligible, we require the radion mass be smaller than the mass of Kaluza--Klein fields \cite{Mishra:2023kiu} 
\begin{equation}\label{CondMass}
    m_{\vf}^2\lesssim 1 \;.
\end{equation}

Below we consider $T_c\simeq 1$ TeV 
to be the benchmark value of the critical temperature, as motivated by the gauge hierarchy problem, but we will also consider values in the range $10^3\:\text{GeV}\lesssim T_c\lesssim 10^{12}\:\text{GeV}$ to probe holographic phase transitions motivated by supersymmetry breaking~\cite{Buyukdag:2018ose,Buyukdag:2018cka,Craig:2020jfv} and/or the Peccei--Quinn mechanism~\cite{Cox:2021lii}.
The parameter $m_\vf^2$ will be varied to probe different regimes of the phase transition, under the constraints imposed by eqs.~\eqref{CondField} and \eqref{CondMass}.
Of particular interest is the regime when $T_q$ is higher than the nucleation temperature $T_n$, i.e. $T_n<T_q<T_c$, in which case the transition occurs via thermally-assisted tunneling.

\section{Euclidean solutions and decay rate}
\label{sec:sol}

Within the approximation of small back-reaction discussed in section~\ref{sec:setup}, the free energy density of the confined phase $F_{\rm RS}$ equals the value, $V_{\rm min}$ of the radion potential (\ref{EffPot}) at the minimum $\vf=\vf_{\rm min}$.
The free energy density can be set to zero by tuning the 4D cosmological constant so that in a constant background $\vf$ we assume\footnote{Note that in \cref{eq:Seff} we factor out the semiclassical parameter $1/N_c^2$ and define the radion potential (\ref{EffPot}) without this factor. }
\begin{equation}\label{Ftrue}
    F_{\rm RS}(\vf)= \frac{3N_c^2}{2\pi^2} \bigl( V(\vf)-V_{\rm min} \bigr) \;.
\end{equation}
Instead, the free energy density of the deconfined phase is obtained by evaluating the 5D gravitational action of the AdS-S solution. It can be written as
\begin{equation}\label{Ffalse}
    F_{\rm AdS-S}(T)=\frac{\pi^2}{8}N_c^2(T_c^4-T^4) \;,
\end{equation}
where $T_c$ is the critical temperature. Moreover, the two free energy densities must coincide in the limit $\vf\to 0$, $T\to 0$, since in this limit they correspond to the same geometry. Thus, from \cref{eq:rad_pot_approx,Ftrue,Ffalse} we obtain 
\begin{equation}\label{Vmin}
    V_{\rm min}=-\frac{\pi^4}{12}T_c^4 \;.
\end{equation}

In the saddle-point approximation, the decay rate (\ref{Gamma}) is obtained by using the bounce solution of the Euclidean equations of motion that interpolates between the two phases.
In the 5D picture, the bounce is a gravitational instanton interpolating between the geometries specified by \cref{BBMetric,RSmetric}.
Constructing this solution explicitly is challenging. Instead, one can resort to the 4D description and work in terms of the radion field $\vf$. One must ensure that the dominant contribution to the Euclidean action of the bounce comes from the region where the 4D theory is valid. One can argue that this is indeed the case as soon as eqs.~\eqref{CondField}, \eqref{CondMass} are satisfied \cite{Creminelli:2001th,Agashe:2019lhy,Agashe:2020lfz,Mishra:2023kiu}.

Under this assumption, the problem reduces to finding the solution of the Euclidean equation of motion for $\vf$ following from the flat-space action (\ref{eq:Seff}).\footnote{
We neglect the spacetime curvature induced by the false vacuum energy density. This is justified as long as the size of the bounce $\ell_b$
is much smaller than the Hubble size in the false vacuum $H\sim T_c^2/(M_PN_c)$. 
Below we will see that $\ell_b\sim T_n^{-1}$, resulting in the bound $T_n\gg T_c^2/(M_PN_c)$ which is always satisfied in our calculations. 
} 
Assuming spherical symmetry with radial coordinate $r=\sqrt{x_1^2+x_2^2+x_3^2}$, the Euclidean equation of motion for the bounce takes the form
\begin{equation}\label{EoM}
    \partial_\tau^2\vf + \partial_r^2\vf + \frac{2}{r}\partial_r \vf- V'(\vf) = 0 \;.
\end{equation}
This must be supplemented with the conditions of periodicity in the imaginary time, $\tau = it$ and regularity at the origin ($r=0$):
\eq{\left.\d_r\vf\right\vert_{r=0} = 0\,.\label{eq:bc_regul}}
Furthermore, we need the boundary condition imposed by the false vacuum. The latter is approached as $\vf\to 0$ but, as discussed in section~\ref{sec:setup}, in this limit the 4D theory (and the potential (\ref{eq:rad_pot_approx})) breaks down.
Still, one can obtain the required condition by matching the Euclidean energy density of the 4D bounce to that of the metastable phase (\ref{Ffalse}) in the region $\vf\approx 0$ \cite{Agashe:2019lhy,Agashe:2020lfz,Mishra:2023kiu}. Since $V(\vf)\approx 0$ at $\vf\approx 0$, the energy density of the bounce in that region is dominated by the gradient term. From \cref{EffPot} and \eqref{Ftrue}-\eqref{Vmin} we obtain
\begin{equation}\label{bc}
\left[ (\partial_\tau\vf)^2 + (\partial_r\vf)^2 \right]_{\vf\approx 0}=\frac{\pi^4}{6} T^4\;.
\end{equation}
This boundary condition, which follows from the continuity of the full 5D potential, causes the solution to be different from the standard 4D thermal bounce solution in flat spacetime.

We numerically solve the equation of motion (\ref{EoM}) with the boundary conditions \eqref{eq:bc_regul} and \eqref{bc}.
First, let us discuss the time-independent configuration that exists at any temperature $T<T_c$. 
The top row of figure~\ref{fig:all_soln} shows this configuration for certain parameters of the radion potential with $m_\vf^2=0.9$ and several values of $T$. The solution agrees with the one studied in Ref.~\cite{Mishra:2023kiu} for
similar 
model parameters. Note that away from $T_c$ the critical bubble cannot be described using the analytic thin-wall approximation and the numerical solution of \cref{EoM} is needed.\footnote{  {At sufficiently low temperatures $T\ll T_c$, when the true vacuum lies
significantly deeper than the potential barrier, the thick-wall approximation may be useful. However, we prefer to use the full numerical bounce solution which is applicable across the full temperature range considered.}}

{
\begin{figure}[t]
    \centering
    \begin{subfigure}[t]{0.06\textwidth}
    \includegraphics[width=\textwidth]{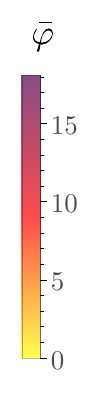}
    \end{subfigure}
     \begin{subfigure}[t]{0.3\textwidth}
    \includegraphics[width=\textwidth]{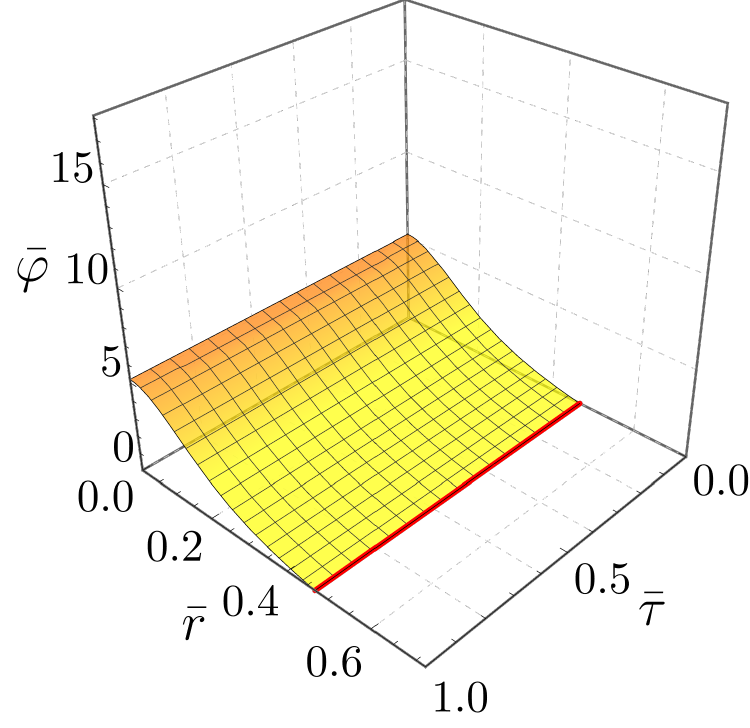}
    \end{subfigure}
    \hfill
    \begin{subfigure}[t]{0.3\textwidth}
    \includegraphics[width=\textwidth]{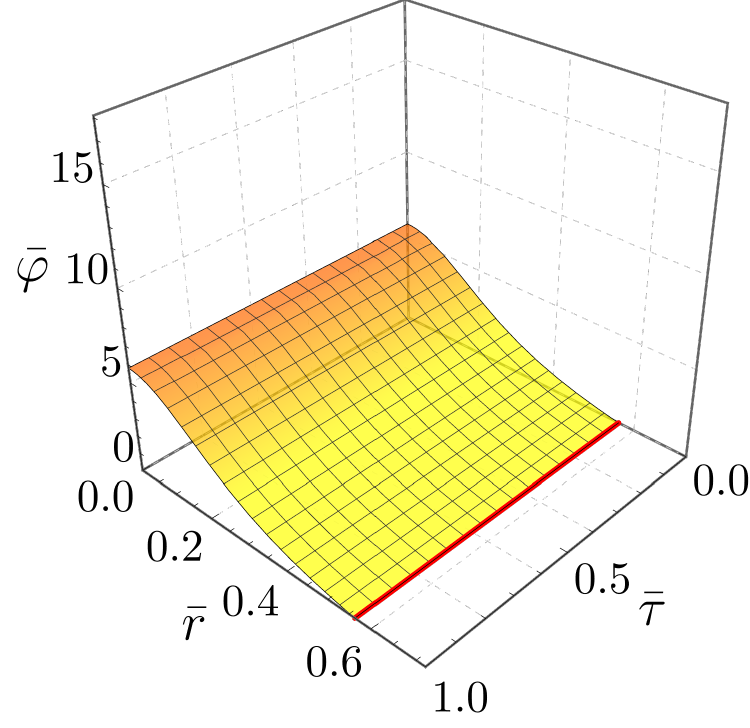}
    \end{subfigure}\hfill
    \begin{subfigure}[t]{0.3\textwidth}
    \includegraphics[width=\textwidth]{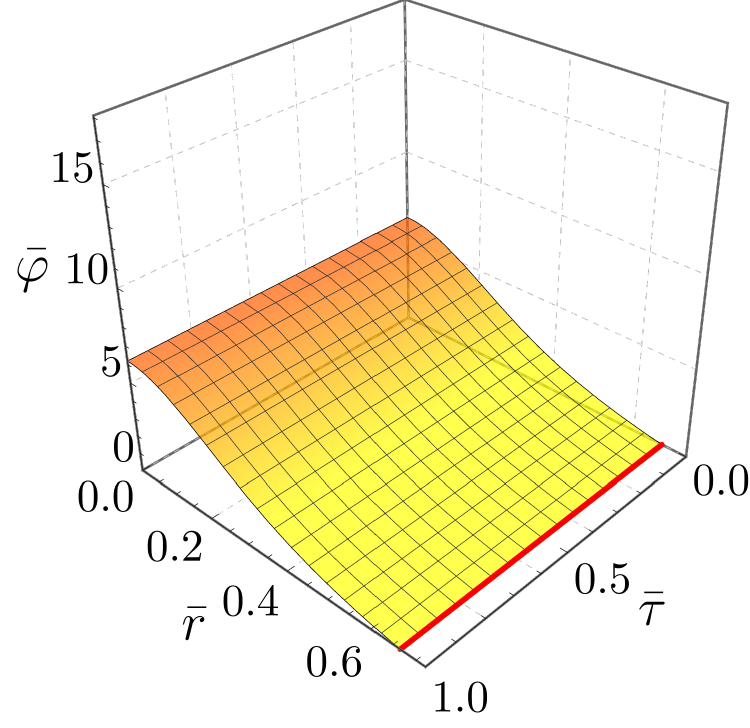}
    \end{subfigure}
   
    \vspace{10pt}
    
    \begin{subfigure}[t]{0.06\textwidth}
    \includegraphics[width=\textwidth]{plots/fig_barlgnd.pdf}
    \end{subfigure}
\hfill
    \begin{subfigure}[t]{0.3\textwidth}
    \includegraphics[width=\textwidth]{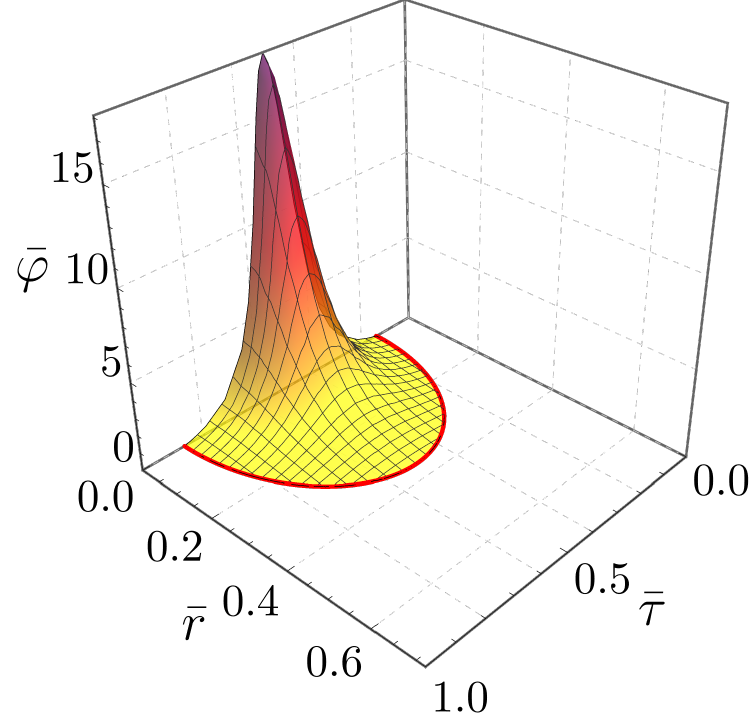}
        \caption{$T = 0.01\, T_c$}\label{fig:O4_bounce}
    \end{subfigure}
    \begin{subfigure}[t]{0.3\textwidth}
    \includegraphics[width=\textwidth]{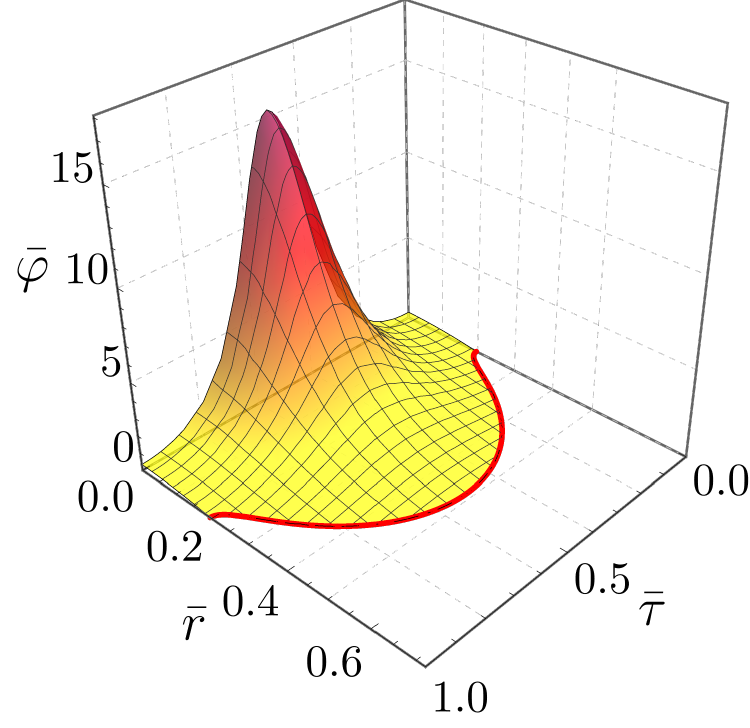}
        \caption{$T = 0.05\, T_c$}\label{fig:bounce_soln_b}
    \end{subfigure}\hfill
        \begin{subfigure}[t]{0.3\textwidth}
    \includegraphics[width=\textwidth]
    {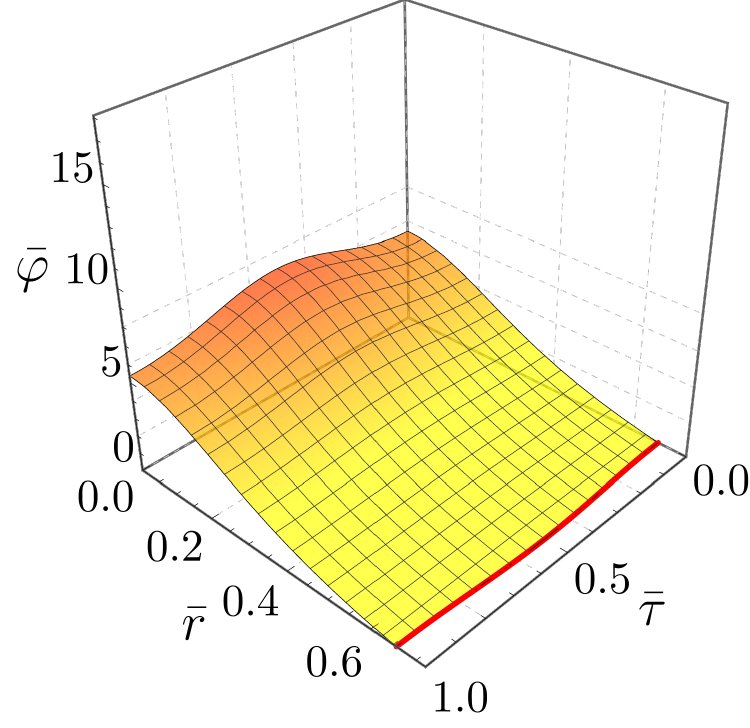}
    \caption{$T \approx T_q = 0.13\, T_c$}\label{fig:bounce_soln_Tq}
    \end{subfigure}
    \caption{ The static critical bubble (top row) and thermal bounce (bottom row) solutions, ${\bvf = \vf_b(\bt,\br)/T}$, 
    in the 4D effective radion theory (\ref{eq:Seff}),
    in the quantum regime (${T<T_q}$), and as a function of the dimensionless coordinates $\bt=\tau T$, $\br=rT$. 
    We assume $\kappa^4=0.1$, $\epsilon_2=-0.04$,  $\epsilon_3 = -0.12, \,\alpha_{\rm IR} = 20$, $ v_{\rm UV} = 0.35$ and $N_c = 3$
 corresponding to $m_\vf^2=0.9$ and $T_c=1$ TeV. (These solutions have explicit dependence only on the two parameters: $T/T_c$ and $m_\vf^2$.) 
    For comparison, both solutions are shown for a single period of $\bt$ and identical ranges of the $\bvf, \br$ axes.
     The boundary where $\vf_b(\bt,\br) = 0$, is shown as a red curve in the $\bt \br$-plane. 
    }
    \label{fig:all_soln}
\end{figure}
}

Next, we study linear perturbations of the critical bubble.
This provides insight into whether there exists another 
time-dependent solution that has a smaller Euclidean action. 
Indeed, the way to understand that the critical bubble is not the relevant solution at $T<T_q$ is by counting the number of its negative eigenmodes -- linearly-independent perturbations that reduce its Euclidean action. The relevant bounce solution for decay must have exactly one such mode \cite{Coleman:1977py,Callan:1977pt}, while the critical bubble at $T<T_q$ has at least two, rendering it irrelevant.

\begin{figure}[t]
    \centering
    \begin{subfigure}[t]{0.49\textwidth}
\includegraphics[width=\linewidth]{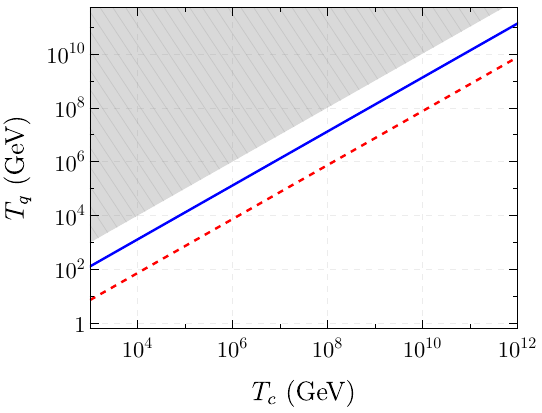}
    \end{subfigure}
       \begin{subfigure}[t]{0.49\textwidth}
\includegraphics[width=\linewidth]{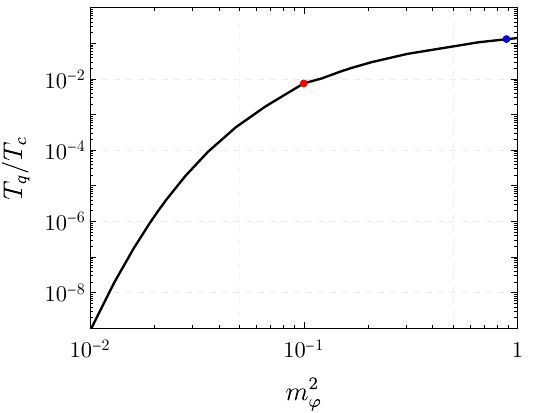}
    \end{subfigure}
    \caption{{\it Left:} The quantum-to-classical transition temperature $T_q$ as a function of the critical temperature $T_c$ for $m_\vf^2 = 0.1$ (dashed red line) and $m_\vf^2 = 0.9$ (solid blue line).
    For $10^3\,{\rm GeV}\leq T_c \leq 10^{12}\,{\rm GeV}$, the ratio $T_q/T_c$ has the constant value of $7.4\times 10^{-3}$ and $0.13$ for $m_\vf^2 = 0.1$ and $0.9$, respectively. The shaded region where $T_q>T_c$ is not physically relevant. {\it Right:} The ratio $T_q/T_c$ as a function of the radion mass-squared $m_\vf^2$ depicting the two benchmark values (solid points) used in our analysis. Note that as $m_\vf^2$ increases, there is more parameter space where the thermal bounce is important.
    }
    \label{fig:Tq}
\end{figure}

The transition temperature $T_q$ at which an extra negative mode appears in the spectrum of perturbations of the critical bubble is determined numerically. The details of this analysis are presented in appendix~\ref{app:modes}. 
We are interested in how $T_q$ behaves as a function of $T_c$ and $m_\vf^2$. Recall that $m_\vf^2$ controls the curvature of the potential (see figure~\ref{fig:rad_pot}) and, hence, has a large effect on $T_q$. This is due to the fact that as $m_\vf^2$ increases, the potential curvature increases, causing the vacuum decay to become more dominated by quantum tunneling.
The dependence of $T_q$ on $T_c$ and $m_\vf^2$ is shown in figure~\ref{fig:Tq}.
We see that for the lowest benchmark value $m_\vf^2=0.1$, the value of $T_q$ is too small for quantum tunneling to play a role in cosmological phase transitions at the TeV scale. 
This agrees with the analysis of Ref.~\cite{Agashe:2019lhy} where, using the same holographic model, it was argued that the classical transitions described by the $O(3)$-symmetric solution dominate over the decays via tunneling, provided the coefficient of the $\vf^4$-term in the radion potential (\ref{eq:rad_pot_approx}) is sufficiently small.
However, the thermal bounce can still become relevant for much larger values of $T_c$, thus allowing for a larger parameter space to be explored.

On the other hand, for $m_\vf^2=0.9$, the quantum-to-classical transition occurs at $T_q\sim 100$ GeV (as seen in figure~\ref{fig:Tq}), which implies quantum tunneling would be relevant for a TeV-scale phase transition. Thus, the thermal bounce dominates vacuum decays as soon as $T_n/T_c\lesssim 10^{-1}$ for all $T_c\gtrsim {\rm TeV}$. This includes the region of parameters explored in Ref.~\cite{Mishra:2023kiu} under the assumption of the $O(3)$-symmetric bounce.

Let us now determine the relevant thermal bounce below the transition temperature. For $T\approx T_q$, we use the critical bubble perturbed along the extra negative mode as an initial guess for the solution. We then decrease the temperature iteratively and construct the bounce at all $T<T_q$. The details of the numerical calculation are presented in appendix~\ref{app:Num}. 
The bottom row of figure~\ref{fig:all_soln} shows the time-dependent bounce solution for the same model parameters and temperature as the static critical bubble in the top row. We see that as $T$ approaches $T_q$, the thermal bounce becomes more identical to the critical bubble solution. 
Instead, for $T\ll T_q$, the solution deviates significantly from the critical bubble and, eventually, the solution $\vf>0$ becomes confined to a region that is less than one period in the imaginary time.
At this point, the bounce becomes  
$O(4)$-symmetric on the interval $0<\tau<1/T$.
Note, however, that the bounce profile is still temperature-dependent via the boundary condition (\ref{bc}). In particular, the size of the bounce scales as $\ell_b\sim T^{-1}$.

Next, we calculate the decay rate given in \cref{Gamma}. The exponent $B(T)$ of the exponential suppression in the decay rate is the Euclidean action of the thermal bounce $\vf_b$ minus the free energy of the false vacuum.
Substituting $\vf_b$ into \eqref{eq:Seff} and using \cref{Ffalse,Vmin}, we obtain
\begin{equation}
\label{BounceAction}
B(T) = \frac{6}{\pi} N_c^2\int\diff\tau\,\diff r ~r^2\left( \frac{1}{2}(\d_\tau\vf_b)^2 + \frac{1}{2}(\d_r\vf_b)^2  + V(\vf_b) + \frac{\pi^4}{12}T^4 \right) \;,
\end{equation}
where the integration is over the region $\vf_b>0$. 
The exponent $B(T)$ is proportional to $N_c^2$ and therefore $1/N_c^{2}\ll 1$ is a small parameter justifying the semiclassical approximation in calculating the decay rate. Furthermore, the temperature dependence in $B(T)$ is only via the ratio $T/T_c$, which can be seen by rescaling the variables $r\to r/T$, $\tau\to\tau/T$, $\vf\to \vf T$ in the action (\ref{eq:Seff}) with the potential (\ref{eq:rad_pot_approx}) and using \cref{Vmin}.

\begin{figure}[t]
    \centering
    \includegraphics[width=0.6\linewidth]{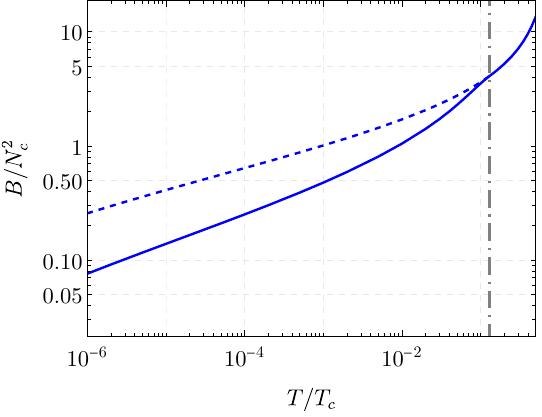}
    \caption{The action $B$ (scaled by $1/N_c^2$) of the dominant false vacuum decay solution (solid blue line) as a function of the temperature $T$ for the large radion mass $m_\vf^2=0.9$.
    Below the temperature $T_q$, depicted by the vertical, dot-dashed gray line, the thermal bounce dominates over the critical bubble solution. For $T<T_q$, the critical bubble action is shown by the blue dashed line for comparison.
    }
    \label{fig:plotB}
\end{figure}

In figure~\ref{fig:plotB} we show how the bounce action $B(T)$ changes when one switches from the critical bubble to the relevant thermal bounce, 
assuming $m_\vf^2=0.9$, which is the phenomenologically relevant case for a TeV-scale phase transition.
We see that, as the temperature decreases below $T_q$, the difference between the two actions can amount to an almost order of magnitude, and therefore will significantly effect the characteristics of the phase transition as will be discussed in the next section. 

The decay rate \eqref{Gamma} also depends on the prefactor $A(T)$. However, the calculation of the prefactor 
is challenging. On dimensional grounds and from zero-mode counting, one can write approximate expressions for $A$ in the classical and quantum regimes of the decay \cite{Linde:1981zj},
\begin{equation}
\label{Prefactor}
    A(T)\approx T^4 \left(\frac{B}{2\pi}\right)^\gamma \;,
\end{equation}
where $\gamma=3/2$ ($\gamma=2$) at $T>T_q$ ($T<T_q$), and we used the fact that for $T\lesssim T_q$ the size of the bounce is $\ell_b\sim T^{-1}$, see figure~\ref{fig:all_soln}. The expression \eqref{Prefactor} will be used to determine the nucleation temperature in section~\ref{sec:transit}.

Before closing this section, let us make an important comment.
The temperature $T_q$, below which decays via thermal tunneling dominate those via classical transitions, has been identified 
with the temperature at which the critical bubble develops an extra negative mode. 
This implies that the thermal bounce describing the tunneling is deformed continuously and merges with the critical bubble in the limit $T\to T_q$, and figure~\ref{fig:all_soln} seems to suggest that this is the case. 
However, a closer examination of the bounce solution near $T_q$ reveals a more complicated behavior, as discussed in appendix~\ref{app:Tq}.
It turns out that the thermal bounces do not branch off from the critical bubble at $T=T_q$,\footnote{
This behavior is suspected from figure~\ref{fig:plotB} by noticing that the action $B(T)$ is not a smooth function of $T$ at $T=T_q$.} nor does $T_q$ exactly equal the temperature at which the second negative mode appears. The difference between the two temperatures is negligible in practice, and amounts to less than one percent for the case $m_\vf^2=0.9$. Nevertheless, it reveals that there are subtleties associated with the holographic bounce that do not typically appear in the usual 4D vacuum decay.

\section{Nucleation temperature }
\label{sec:transit}

At high temperatures $T> T_c$, the universe begins in the deconfined phase and eventually cools to $T_c$ where the vacuum of the confined phase becomes degenerate with the deconfined phase.
The phase transition begins when the decay rate of the deconfined phase becomes comparable with the Hubble expansion rate. This process nucleates bubbles of the new confined phase at the nucleation temperature $T_n$. Since there is a large potential barrier separating the phases, the phase transition is delayed and occurs with strong supercooling where $T_n/T_c\ll 1$. As the temperature decreases from $T_c$ to $T_n$, the universe therefore remains in the false vacuum  causing a brief period of inflation where the radiation is subsequently diluted. 
The Hubble rate is approximately constant and is determined by the energy density $\rho_{BB}$ of the black brane:
\begin{equation}\label{Hubble}
    H^2 = \frac{\rho_{BB}}{3M_P^2} \;, \qquad  \rho_{BB}=\frac{8\pi^2}{N_c^2}T_c^4 \;.
\end{equation}
Equating $H^4$ to the decay rate (\ref{Gamma}), we obtain the equation for the nucleation temperature which we write in the form
\begin{equation}\label{T_nucl}
       B\left( \frac{T_n}{T_c} \right) - 4\log\frac{T_n}{T_c} =  2\log\left(\alpha\frac{ M_P^2 N_c^{2}}{T_c^2} \right) \;,
\end{equation}
where $\alpha=\frac{3}{8\pi^2}\left(\frac{B}{2\pi}\right)^{\gamma/2}$ contains the subleading dependence on $T_n$.
We look for the solution to \eqref{T_nucl}  
at a given $T_c$, $N_c$ and $m_\vf^2$ subject to the constraints $N_c\geqslant 3$ and $B\gg 1$ required by the semiclassical approximation and exponential suppression of the decay rate.
We are interested in the regime $T_n<T_q<T_c$ when thermal tunneling dominates the phase transition. At the same time, the nucleation temperature should not be too low, $T_n\gtrsim 10$ MeV, since we do not want to alter the successful predictions of big bang nucleosynthesis \cite{ParticleDataGroup:2024cfk}.
Importantly, the solution of \cref{T_nucl} may not exist at sufficiently high values of $T_c$ and $N_c$, indicating that the phase transition is not always possible. This is due to the fact that $\Gamma(T)$ has a maximum at some finite $T$ (within the validity of the semiclassical approximation). 

\begin{figure}[t]
\centering
       \begin{subfigure}[t]{0.49\textwidth}
\includegraphics[width=\linewidth]{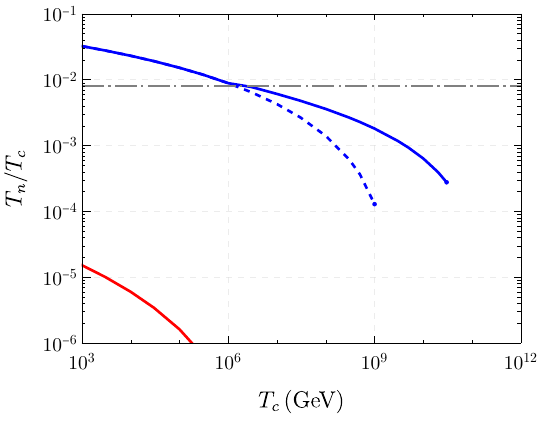}
    \end{subfigure}
\begin{subfigure}[t]{0.49\textwidth}
\includegraphics[width=\linewidth]{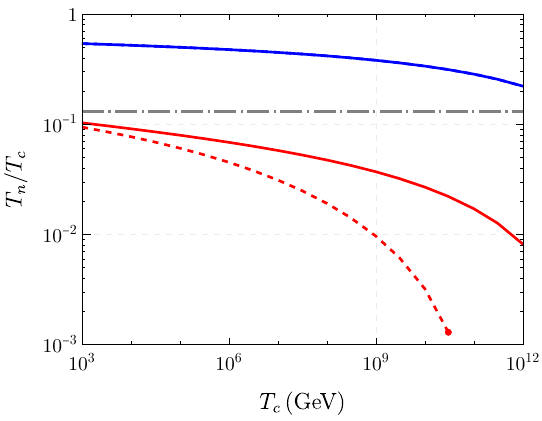}
    \end{subfigure}
\caption{The nucleation temperature $T_n$ (scaled by 1/$T_c$) as a function of the critical temperature $T_c$ for $N_c = 3$ ($6$) shown as solid blue (red) lines,
assuming $m_\vf^2 = 0.1$ (left plot) and $0.9$ (right plot). 
For comparison, the nucleation temperature determined from the critical bubble solution is shown as a dashed blue (red) line for $N_c = 3$ ($6$). 
The horizontal gray dot-dashed line indicates the classical-to-quantum transition temperature, $T_q$ (again scaled by $1/T_c$) which separates the classical and quantum regimes. The values of the ratio where $T_n/T_c>1$ are not physically relevant. Note that for each value of $N_c$, the curves end at a maximum value of $T_c$, above which the transition never takes place $(i.e.~\Gamma < H^4)$. 
}
    \label{fig:Tn}
\end{figure}

We find that for small radion mass squared, $m_\vf^2=0.1$, corresponding to the absence of nonlinearity in the Goldberger--Wise potential (i.e. $\epsilon_3=0$), the above conditions cannot be satisfied for a TeV-scale transition with $N_c =3$, and the critical bubble is the relevant solution. 
This is illustrated on the left panel of figure~\ref{fig:Tn} and agrees with the analysis of Ref.~\cite{Agashe:2020lfz}. On the contrary, for $N_c = 6$, the thermal bounce solution crucially enhances the vacuum decay rate to complete the phase transition at the TeV scale, which would not be possible using only the critical bubble solution.
Moreover, the thermal bounce remains relevant at higher critical temperatures, $T_c\gtrsim 10^6$ GeV for $N_c=3$.
Similarly, for large radion mass, $m_\vf^2=0.9$, the required conditions can be satisfied already for the TeV scale transition with $N_c =6$, as is shown on the right panel of figure~\ref{fig:Tn}.
It is then important to use the thermal bounce action (\ref{BounceAction}) to find the nucleation temperature correctly.
For example, using the critical bubble instead of the thermal bounce would lead to underestimating $T_n$ by one order of magnitude at $T_c=10^{10}$ GeV and $N_c=6$.
 {To get the correct value of $T_n$ using the critical bubble, one would have to shift the radion mass by a factor of $\mathcal{O}(1)$.}
Note that the endpoints of the functions $T_n(T_c)$ shown in figure~\ref{fig:Tn} correspond to the maximum of the decay rate, beyond which $\Gamma(T)$ is always smaller than $H^4$ and no phase transition occurs.
\begin{figure}[t]
\centering
       \begin{subfigure}[t]{0.49\textwidth}
\includegraphics[width=\linewidth]{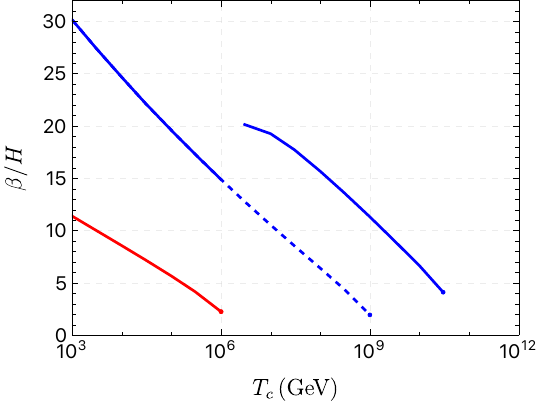}
    \end{subfigure}
\begin{subfigure}[t]{0.49\textwidth}
\includegraphics[width=\linewidth]{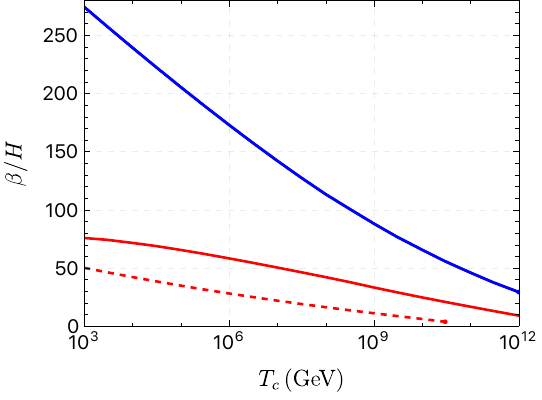}
    \end{subfigure}
\caption{{The ratio $\beta/H$, defined in \eqref{eq:betaH}, as a function of the critical temperature $T_c$ for $N_c = 3$ ($6$) shown as solid blue (red) lines,
assuming $m_\vf^2 = 0.1$ (left plot) and $0.9$ (right plot). 
For comparison, the value of $\beta/H$ determined from the subdominant critical bubble solution is shown as a dashed blue (red) line for $N_c = 3$ ($6$). In the left plot, the solid blue line exhibits a discontinuity near $T_c = 10^6$ GeV, which corresponds to where the thermal bounce branches off from the critical bubble (see figure~\ref{fig:plotB}). 
}
}
    \label{fig:betaH}
\end{figure}
\section{Gravitational wave signal}
\label{sec:pheno}

In this section, we calculate the GW signal that arises from the phase transition mediated by the quantum thermal bounce. We assume that the dominant source of the gravitational waves is due to the collision of bubble walls. This is reasonable, since at large supercooling the ambient radiation is dilute and exerts little pressure on the bubbles which, therefore, can expand in the run-away mode \cite{Caprini:2015zlo,Caprini:2019egz}. Upon completion of the phase transition, most of the energy contained in the bubble walls is transferred to the thermal plasma, which attains a temperature $T_*\approx T_c$ {; this is a good approximation for a strong first-order transition where the large latent heat reheats the universe}.

\begin{figure}[t]
    \centering
    \includegraphics[width=.9\linewidth]{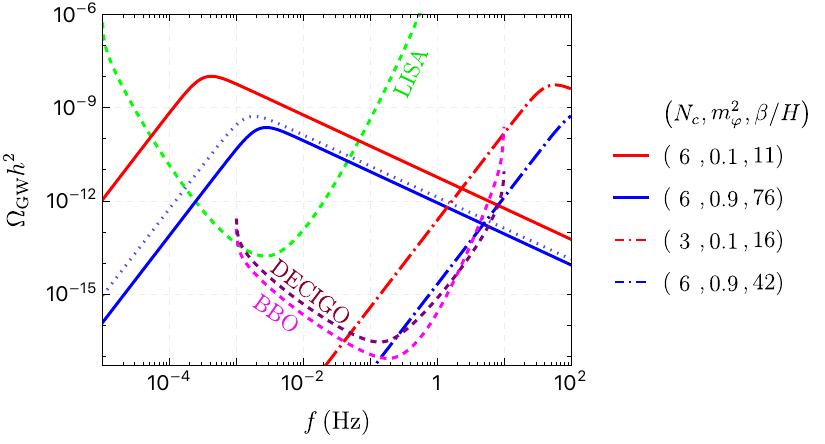}
    \caption{
    The  {fractional energy density} of the stochastic GWs as a function of frequency due to collisions of bubbles nucleated via thermal tunneling. The red (blue) lines correspond to $m_\vf^2 = 0.1$ ($0.9$), assuming $g_* = 100$ and $T_c = T_* = 10^3$ GeV (solid lines) and $10^8$ GeV (dot-dashed lines).  {The blue dotted line shows the GW signal as predicted by the static critical bubble solution, which compares with the true time-dependent bounce solution (blue solid line) for the same parameters.}
    The green, magenta and purple dashed lines show the projected discovery reach from LISA \cite{LISA:2017pwj,Baker:2019nia}, BBO \cite{Crowder:2005nr,Corbin:2005ny,Harry:2006fi,Yagi:2011wg} and DECIGO \cite{Seto:2001qf,Kawamura:2006up,Isoyama:2018rjb}, respectively.
    }
    \label{fig:GWsig}
\end{figure}

The GW signal is characterised by the peak frequency $f_p$ and the frequency-dependent abundance $\Omega_{\rm GW}h^2$. For the GWs generated by bubble collisions these quantities are given by \cite{Caprini:2015zlo,Caprini:2019egz}
\begin{equation}\label{f_peak}
f_p=0.037\:\text{mHz}\:\frac{\beta}{H}\left(\frac{T_*}{\text{TeV}}\right)\left(\frac{g_*}{100}\right)^{\frac{1}{6}} \;,
\end{equation}
and
\begin{equation}\label{Omega}
\Omega_{\rm GW}h^2(f)=1.3\times 10^{-6}\left( \frac{H}{\beta}\right)^2\left(\frac{100}{g_*}\right)^{\frac{1}{3}}\frac{3.8(f/f_p)^{2.8}}{1+2.8(f/f_p)^{3.8}} \;,
\end{equation}
where $H$ is the Hubble rate during the phase transition, $g_*$ is the number of relativistic degrees of freedom, and $\beta$ is related to the inverse duration of the transition. 
The ratio $\beta/H$ is obtained from the bounce action as \cite{Turner:1992tz}
\begin{equation}
\frac{\beta}{H}=-\left.\frac{\diff\log\Gamma}{\diff\log T}\right\vert_{T=T_n} \;. \label{eq:betaH}
\end{equation}
{In Fig.~\ref{fig:betaH}, we show $\beta/H$ as a function of the critical temperature $T_c$ obtained using our quantum thermal bounce, and for $T_n < T_q$, the $\beta/H$ values are compared with those obtained from the (inaccurate) critical bubble solution.}

We now have all the ingredients to compute the GW signal from the phase transition.
We focus on the parameter space where bubble nucleation occurs via thermally-assisted tunneling. 
As discussed in section~\ref{sec:setup}, this can be easily realized in a model with a nonlinear Goldberger--Wise potential. Allowing for the nonlinearity, we can explore a large range of critical temperatures, $10^3\,{\rm GeV}\lesssim T_c\lesssim 10^{12}\,{\rm GeV}$ that could correspond to phase transitions associated with the electroweak hierarchy, supersymmetry breaking or Peccei--Quinn symmetry breaking.

The GW signal produced by phase transitions dominated by thermally-assisted tunneling is shown in figure~\ref{fig:GWsig}.
As the figure depicts, thermal tunneling does induce a potentially observable signal at future gravitational wave observatories both for the small and large radion mass cases. The small radion mass case ($m_\vf^2=0.1$) requires a larger (smaller) number of colors $N_c$ at lower (higher) values of $T_c$, whereas the larger radion mass case ($m_\vf^2=0.9$) entails larger $N_c$ at any $T_c$.
It is particularly interesting that even for high critical temperatures, $T_c=10^8$ GeV, the signal is within reach of future GW observatories where supercooled phase transitions dominated by tunneling can be probed. Importantly, using only the critical bubble solution would not lead to any GW signal for the TeV-scale phase transition shown in figure~\ref{fig:GWsig} with $N_c=6$ and $m_\vf^2=0.1$.  {Similarly, for larger radion mass $m_\vf^2 = 0.9$, again with $N_c = 6$, the GW signal predicted by the thermal tunneling solution (blue solid line) differs by an order of magnitude from that of the critical bubble solution (blue dotted line)\footnote{ {However, note that readjusting the radion mass by an $\mathcal{O}(1)$ factor can reproduce the same signal obtained from the (inaccurate) critical bubble solution.}}.} Finally, note that while we only showed signals for the benchmark values in figure~\ref{fig:GWsig}, the thermal bounce would also give new signals
for other parameter values, thereby extending the parameter space in holographic models beyond the range studied with the critical bubble.

\section{Conclusion}
\label{sec:concl}

Holographic models provide a novel way to study first-order phase transitions between strongly-coupled phases in finite temperature Yang--Mills theories. The corresponding holographic phase transitions have only been analyzed using the $O(3)$-symmetric (or critical bubble) configurations that describe vacuum decay via classical thermal fluctuations at high temperatures, while at low temperatures the $O(4)$-symmetric (vacuum) bubble has been used.
However, these solutions do not provide an accurate description of phase transitions that occur via thermally-assisted quantum tunneling, corresponding to the thermal bounce -- a time-dependent, periodic Euclidean solution.

In this work, we studied the thermal bounce in the holographic framework which describes the transition between the deconfined and confined phases at strong coupling. In particular, we employed a holographic model with a Goldberger--Wise scalar field to stabilize the location of the IR brane and considered a nonlinear bulk potential with a cubic term that enhances the conformal symmetry breaking in the IR. Working within the 4D effective theory of the radion field, we calculated the thermal bounce numerically and demonstrated how it interpolates between the $O(3)$-symmetric critical bubble at high temperature and $O(4)$-symmetric configuration at low temperature. This is the first time such a solution has been obtained in the holographic framework.

Our result shows that using the $O(4)$-symmetric configuration gives the parametrically correct estimation of the thermal bounce action. One may expect this to be the case from the standard picture of vacuum decay in flat space, where the thermal bounce indeed interpolates between the $O(4)$-symmetric zero-temperature bubble and the critical bubble. In the holographic setup, however, this is a nontrivial statement, especially in models where the limit $T\to 0$ is not well-defined and hence the $O(4)$-symmetric configuration has not been shown to exist.

As an application of our new thermal bounce solution, we calculated the nucleation temperature and the ensuing stochastic GW background in the parameter region where the thermal bounce dominates the decays, therefore complementing the analyses performed in Refs.~\cite{Agashe:2020lfz,Mishra:2023kiu}.
In particular, we explored a large range of the critical temperature, $10^3\:\text{GeV}\lesssim T_c\lesssim 10^{12}\:\text{GeV}$. We showed that even for $T_c\simeq 10^{8}$ GeV one can obtain a potentially observable GW signal if supercooling is sufficiently strong.

Importantly, our solution is calculated in the region of validity of the 4D theory. The full 5D solution -- a gravitational instanton interpolating between the two geometries -- can, in principle, be constructed, but this is challenging even in the $O(3)$-symmetric case \cite{Mishra:2023kiu,Mishra:2024ehr}. Besides, the relevance of such an instanton for vacuum decay is hard to establish, since the negative mode counting argument cannot be literally applied, see \cite{Bramberger:2019mkv} and references therein. Nevertheless, it is an  {important} direction of future research,
 {since it will enable a more accurate determination of the bounce action.}

Furthermore, knowing the full 5D solution is important for an accurate prediction of the bounce action and the decay rate \cite{Mishra:2023kiu,Mishra:2024ehr}.
On the other hand, determining the decay rate prefactor in the thermal holographic framework is even more difficult.
In addition, dynamical effects at the onset of the phase transition can produce large corrections to the thermal formula \cite{Pirvu:2024ova,Pirvu:2024nbe}.
In this case, real-time simulations of phase transitions in strongly-coupled systems by means of holographic duality could help to improve our understanding.

\section*{Acknowledgments}

We thank Junwu Huang, Rashmish Mishra and Sergey Sibiryakov for helpful discussions.
The work of T.G. and A.P. is supported in part by the Department of Energy under Grant No.~DE-SC0011842 at the University of Minnesota. Part of the work of A.P. is supported by the Mikhail Voloshin fellowship. The research of A.S. at Perimeter Institute is supported in part by the Government of Canada through the Department of Innovation, Science and Economic Development Canada and by the Province of Ontario through the Ministry of Colleges and Universities.

\appendix

\section{Negative modes of the critical bubble}
\label{app:modes}

In this appendix, we study the linear perturbations around the critical bubble. Our goal is to count the number of negative eigenmodes around the static solution, i.e. the number of linearly-independent perturbations that reduce its Euclidean action. When more than one such mode is present in the spectrum, the solution is no longer relevant for vacuum decay. Thus, the linear analysis allows us to find the parameter region where the time-dependent thermal bounce may exist. Moreover, we can use the critical bubble perturbed along its extra negative mode as an initial guess in the numerical search of the thermal bounce, as explained in appendix~\ref{app:Num}.

It is convenient to work with 
the dimensionless variables $\bvf, \br, \bt$ and the rescaled potential $\bV(\bvf, T)$ defined as
\eq{
\bar{\vf} = \vf / T\,,~
\br =  r\, T\,,~\bt =  \tau \, T\,,~ \bV &= \frac{1}{T^4} V\left(T\, \bvf\right).\label{eq:def_dimless}
}
Note that $\bV(\bvf, T)$ explicitly depends on the temperature due to the conformal symmetry breaking term in the potential \eqref{EffPot}. The equation of motion \eqref{EoM} and the boundary conditions \eqref{eq:bc_regul}, \eqref{bc} become
\eq{
&\frac{\partial^2 \bvf}{\partial \bt^2} + \frac{\partial^2 \bvf}{\partial \br^2} + \frac{2}{\br} \frac{\partial \bvf}{\partial \br} = \frac{d\bV}{d\bvf}\,,\label{eq:eom2}\\
&\left.\frac{\partial \bvf}{\partial \br}\right\vert_{\br = 0} = 0 \,,\qquad\left[\left(\frac{\partial \bvf}{\partial \bt}\right)^2+\left(\frac{\partial \bvf}{\partial \br}\right)^2\right]_{\bvf\approx0} = \frac{\pi^4}{6}\label{eq:bc2}\,.
}
Notice that the boundary condition \eqref{eq:bc2} at $\bvf \approx 0$ is temperature independent unlike \eqref{bc}.
Since the critical bubble $\bvf=\bvf_0(\br)$ is time independent, 
the eigenmodes $\bar{\chi}_I(\bt,\br)$ of the linear perturbations around $\bvf_0(\br)$ can be found using the separation of variables, 
\begin{equation}
    \bar{\chi}_I(\bt,\br)=\bar{\chi}_{n,k}(\br)\e^{-i\bar{\omega}_n\bt} \;,
\end{equation}
where the index $I$ enumerates the eigenmodes and  $\bar{\omega}_n=2\pi n$ are the rescaled Matsubara frequencies.
The spatial functions $\bar{\chi}_{n,k}(\br)$ satisfy the differential equation 
\begin{equation}\label{eq:evalue_eqn1}
    -\bar{\chi}_{n,k}'' - \frac{2}{\br}\,\bar{\chi}_{n,k}' + \left[\bar{\omega}_n^2 + \frac{\diff^2\bar{V}(\bar{\vf}_0)}{\diff\bar{\vf}^2} \right]\bar{\chi}_{n,k} = \lambda^{(k)}_n \bar{\chi}_{n,k} \,,
\end{equation}
where $\lambda^{(k)}_n$ are the eigenvalues and $^\prime$ denotes the derivative with respect to $\br$.
The boundary conditions for the eigenmodes following from \cref{eq:bc2} take the form
\begin{equation}\label{bc_modes}
    \left.\bar{\chi}'_{n,k}\right\vert_{\bar{r}=0}=0 \;, ~~~ \left.\frac{\bar{\chi}'_{n,k}}{\bar{\chi}_{n,k}}\right\vert_{\br=\br_0} = -\frac{2}{\br_0} \;,
\end{equation}
where $\br_0$  is the location of the boundary that arises 
from the applicability of the 4D theory, i.e. $\bvf_0(\br_0) = 0$. The differential equation (\ref{eq:evalue_eqn1}) can be further simplified by the change of variable, $ \bar{\chi}_{n,k}(\br) \equiv \frac{u_k(\br)}{\br}$, which leads to
\begin{align}
& -u''_k(\br) + W(\br)\,u_k(\br) = \Lambda^{(k)} u_k(\br)\,, \label{eq:evalue_eqn3} \\
& ~~~~u_k(0) = 0 \;, ~~~ \left.\frac{u'_k}{u_k}\right\vert_{\br=\br_0}= -\frac{1}{\br_0} \;, \label{bc_modes2}
\end{align}
where $W(\br)\equiv \frac{\diff^2\bar{V}(\bar{\vf}_0)}{\diff\bar{\vf}^2}$ and $\Lambda^{(k)}\equiv\lambda^{(k)}_n-\bar{\omega}_n^2$.\footnote{The index $n$ in the definitions of $u_k$ and $\Lambda^{(k)}$ has been dropped, since in \cref{eq:evalue_eqn3} there is no explicit dependence on $n$.}

To obtain an intuition for the eigenvalue spectrum that results from solving \cref{eq:evalue_eqn3}, we note that \eqref{eq:evalue_eqn3} mimics an eigenvalue problem corresponding to the Hamiltonian operator in one-dimensional quantum mechanics with the potential $W(\br)$ and the energy eigenvalue $\Lambda^{(k)}$.
Since $\bar{V}(\bvf)$ is concave down for $\bvf\ll \bvf_{\rm min}$, we have $W(\br)<0$. Hence, there is at least one negative eigenvalue $\Lambda^{(0)}=\lambda_0^{(0)}$ corresponding to the ground state in the potential $W(\br)$. This agrees with the fact that the critical bubble, being a saddle point, should have at least one negative direction in the configuration space. 
Provided that the ground state is well localized in the region $\br<\br_0$, we expect that the boundary condition (\ref{bc_modes2}) does not change this reasoning.

\begin{figure}[t]
\begin{subfigure}[t]{.49\linewidth}
\includegraphics[width=\textwidth]{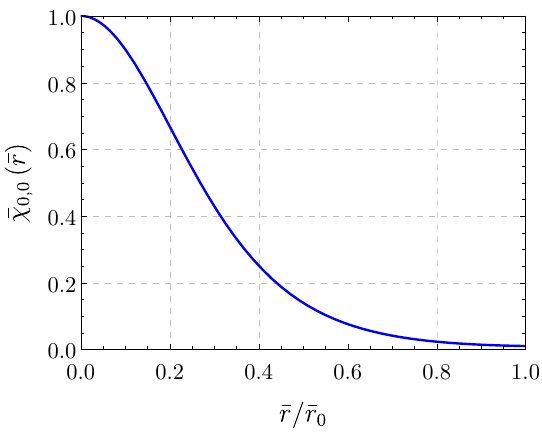}
       \end{subfigure}
              \hfill
       \begin{subfigure}[t]{.49\linewidth}
           \includegraphics[width=\textwidth]{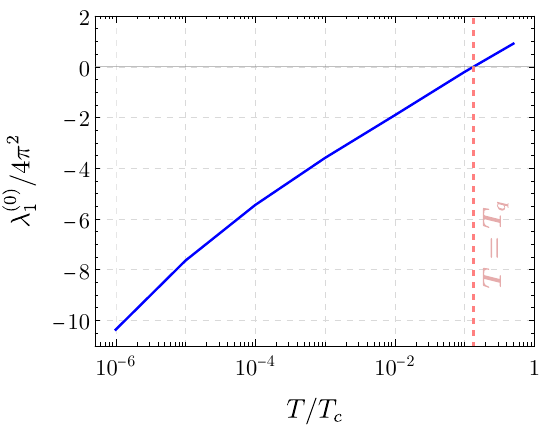}
       \end{subfigure}
       \caption{{\it Left:} The eigenstate $ \bar{\chi}_{0,0}(\br)$ as a function of $\br/\br_0$, assuming a temperature $T = 0.01T_c$ and $m_\vf^2=0.9$.
       {\it Right:} The eigenvalue $\lambda^{(0)}_1$ obtained from solving \eqref{eq:evalue_eqn1}, as a function of $T/T_c$ assuming $m_\vf^2=0.9$.  The vertical, dashed pink line 
       depicts the temperature $T_q$ below which a new negative mode appears. 
       } \label{fig:evalue_epsilon3}
\end{figure}

To verify this intuition, we solve numerically \cref{eq:evalue_eqn3} at different temperatures.
At any $T$ we find the ground state $\bar{\chi}_{0,0}(\br)$ and the corresponding eigenvalue $\lambda_0^{(0)}$. As shown in figure~\ref{fig:evalue_epsilon3}, the function $\bar{\chi}_{0,0}(\br)$ becomes exponentially small at $\br\approx\br_0$, and therefore the boundary condition (\ref{bc_modes2}) does not modify the spectrum. Furthermore, we find that the value $\lambda_0^{(0)}$ decreases as the temperature decreases. As soon as it drops below $-4\pi^2$, the second negative eigenvalue appears in the dipole sector $n=1$:
\begin{equation}
   \l_0^{(0)} = \Lambda^{(0)} < -4\pi^2 ~~~ \Rightarrow  ~~~ \l_1^{(0)}=\Lambda^{(0)}+\bar{\omega}_1^2 < 0 \;.
\end{equation}
The eigenvalue $\l_1^{(0)}$ is shown in figure~\ref{fig:evalue_epsilon3} as a function of $T/T_c$. We see that it indeed becomes negative at a certain temperature $T_q$, indicating that at $T<T_q$ the critical bubble is no longer the relevant solution, and one must look for the time-dependent thermal bounce.

\section{Numerical calculation of the thermal bounce}
\label{app:Num}

In this appendix, we outline the numerical procedure used to determine the thermal bounce solution and its Euclidean action. Working in the coordinates $(\bt,\br)$ defined in \cref{eq:def_dimless}, we solve the equation of motion (\ref{eq:eom2}) with the boundary conditions (\ref{eq:bc2}) as well as the condition of periodicity in the imaginary time.
Note that the location of the thermal bounce boundary $\bvf\approx 0$ is not a priori known in the $\bt \br$-plane. 
Assuming that the boundary is located at 
\eq{\bar{r}=\br_b(\bt) \equiv \e^{-\zeta(\bt)}\,,\label{eq:bdry}} 
we must solve for the function $\zeta(\bt)$. 
For the time-dependent solution, the boundary is a curve in the $\bt \br$-plane, which is difficult to treat numerically. 
Therefore, we map the integration domain into a rectangle by the following transformation of coordinates:
\eq{ \tr = \br \e^{\zeta(\bt)}\,,~~\tdt = \bar{\tau}\,.\label{eq:tildecoords}}
In the coordinates ($\tdt,\tr$), the integration domain now becomes
$0\leqslant\tilde{\tau}<1$ and $0\leqslant\tilde{r}\leqslant 1$.
The $\tdt$ range corresponds to the period $1/T$ in the imaginary time $\tau$, and the $\tr$ range follows from the boundary location (\ref{eq:bdry}).
The equation of motion \eqref{eq:eom2} and the boundary conditions \eqref{eq:bc2} in the coordinates \eqref{eq:tildecoords} are given by
\begin{align}
 & \frac{\partial^2 \bvf}{\partial \tdt^2} +2\tr\dot\zeta\frac{\partial^2 \bvf}{\partial \tdt \partial \tr} +\left(\e^{2\zeta}+\tr^2\dot\zeta{}^2\right) \frac{\partial^2 \bvf}{\partial \tr^2} +\left(\tr\ddot\zeta+\tr\dot\zeta{}^2+\frac{2}{\tr} \e^{2\zeta}\right)  \frac{\partial \bvf}{\partial \tr} =  \frac{d\bV}{d\bvf} \,, \label{eq:eomNew}  \\
 &  \left.\frac{\partial \bvf}{\partial \tr}\right\vert_{\tr=0} = 0\,,~~~\left(\e^{2\zeta}+ \dot\zeta{}^2\right)\left.\left(\frac{\partial \bvf}{\partial \tr}\right)^2\right\vert_{\tr=1} = \frac{\pi^4}{6}\,,\label{eq:bcNew}
\end{align}
where dot denotes the derivative with respect to $\tdt$.
In addition, a Dirichlet boundary condition is imposed on $\bvf$ at the boundary $\tr=1$, namely
\begin{align}
\bvf(\tilde{\tau},1)=0\,.
    \label{eq:bc_Drchlt}
\end{align}

\begin{figure}[t]
  \begin{subfigure}[t]{0.4 \linewidth}
  \centering
      \includegraphics[width=.99\linewidth]{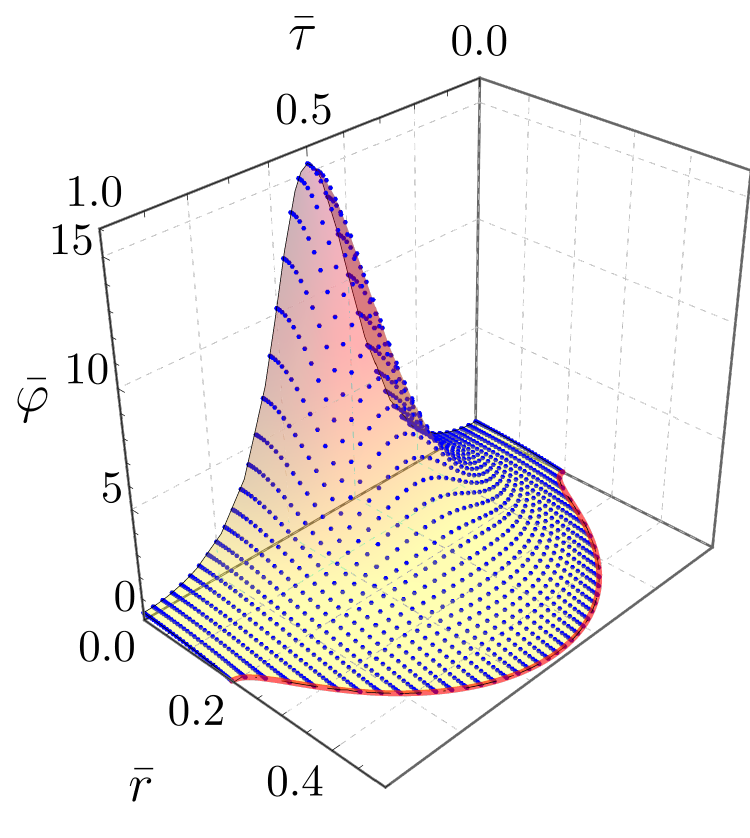}
  \end{subfigure} \hfill
  \begin{subfigure}[t]{0.6 \linewidth}
  \centering
      \includegraphics[width=.99\linewidth]{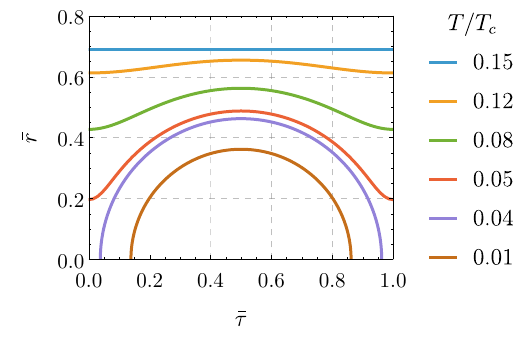}
  \end{subfigure}
    \caption{{\it Left:} The thermal bounce solution $\bvf=\bvf_b(\bt,\br)$ and the corresponding grid (indicated by blue dots above the $\bar{\tau} \bar{r}$-plane) used for the Newton--Raphson method, assuming $T/T_c = 0.05$ and $m_\vf^2=0.9$. 
    {\it Right:} The contours of the boundary ($\varphi \approx 0$) separating the two phases probed by the thermal bounce, in the $\bar{\tau} \bar{r}$-plane and for different values of $T/T_c$. For comparison, the critical bubble wall at $T = 0.15T_c \gtrsim T_q$ is shown as the blue horizontal line.
   } \label{fig:bubble_bdry}
    
\end{figure}

Using the rectangular grid in the $(\tdt,\tr)$-coordinates, we solve \cref{eq:eomNew} with the boundary conditions (\ref{eq:bcNew})-(\ref{eq:bc_Drchlt}) and the condition of periodicity in $\td{\tau}$ by the Newton--Raphson method \cite{Press:2007ipz} to obtain the thermal bounce $\bvf_b(\tdt,\tr)$ and its boundary determined by the function $\zeta(\td{\tau})$.
As an initial guess for the Newton--Raphson method, at temperatures
$T\lesssim T_q$ we use the critical bubble, which can be easily found by solving the corresponding differential equation, which is perturbed around its extra negative mode (see appendix~\ref{app:modes}). We then obtain the solution iteratively by lowering the temperature and using the bounce solution found at the slightly higher temperature as the new initial guess. The solution evaluated at a particular temperature and an example of the grid used to obtain it are shown in figure~\ref{fig:bubble_bdry}.
Finally, using the variables defined in (\ref{eq:def_dimless}) and the coordinates $(\tdt,\tr)$, the bounce action (\ref{BounceAction}) becomes
\begin{equation}
\begin{split}
    B = \frac{6}{\pi} N_c^2\int\displaylimits_0^{1}\diff\tdt\int\displaylimits_0^1 \diff\tr\, \tr^2 \,\e^{-3\zeta}\Bigg[ & \frac{1}{2}\left(\e^{2\zeta} + \tr^2 \dot\zeta{}^2\right)\left(\frac{\partial \bvf_b}{\partial \tr}\right)^2 +\frac{1}{2} \left(\frac{\partial \bvf_b}{\partial \td{\tau}}\right)^2  \\ 
    &  + \tr\dot\zeta \frac{\partial \bvf_b}{\partial \td{\tau}} \frac{\partial \bvf_b}{\partial \tr}+ \bV(\bvf_b) + \frac{\pi^4}{12}\Bigg]\,. \label{eq:Bth}
\end{split}
\end{equation}

Note that at sufficiently low temperatures the thermal bounce becomes $O(4)$-symmetric on the time interval $0\leqslant \tilde{\tau}<1$ and the coordinates $(\tilde{\tau},\tilde{r})$ become inappropriate. 
This is depicted in the right panel of figure~\ref{fig:bubble_bdry} which shows that as the temperature decreases, the boundary changes from being located at a constant $\br$ value, corresponding to the critical bubble, to being located on the semi-circle $\bt^2 + \br^2=\,$constant, corresponding to the $O(4)$-symmetric solution. Thus, for sufficiently low temperatures it is more appropriate to switch to the radial coordinate $\bar{R} = \sqrt{\bt^2 + \br^2}$. 
The equation of motion (\ref{eq:eom2}) then reduces to an ordinary differential equation, and the boundary value problem (\ref{eq:eom2}), (\ref{eq:bc2}) is straightforward to solve.
The bounce action reads
\eq{B =3N_c^2\int\displaylimits_0^{\bar{R}_b} \diff \bar{R}~\bar{R}^3 \left[\frac{1}{2}\left(\frac{d \bvf_b}{d \bar{R}}\right)^2 + \bV(\bvf_b) +  \frac{\pi^4}{12}\right] \,,}
where $\bar{R}_b$  is the location of the $O(4)$-symmetric boundary, i.e. $\bvf_b=0$.

\section{Thermal bounce near the transition temperature}
\label{app:Tq}

\begin{figure}[t]
    \centering
    \begin{subfigure}{0.45\linewidth}
    \includegraphics[width=\linewidth]{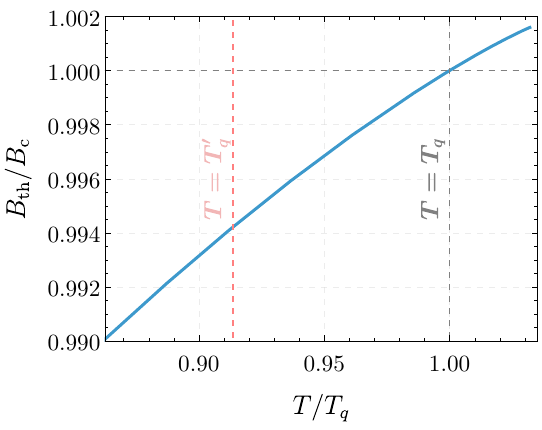}    
    \end{subfigure}
    \hfill
    \begin{subfigure}{0.53\linewidth}
    \includegraphics[width=\linewidth]{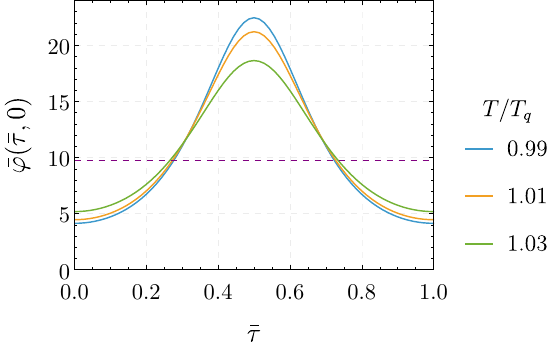}    
    \end{subfigure}
    \caption{\textit{Left:} The ratio of the thermal bounce action $B_{\rm th}$ to the critical bubble action $B_c$ as a function of temperature $T$ near $T_q$, assuming $m_\vf^2=0.1$. The temperature $T_q'$, at which the extra negative mode of the critical bubble appears, is indicated by the pink dashed vertical line.
\textit{Right:} The thermal bounce profile $\bar{\varphi}(\bar{\tau}, \bar{r} = 0)$ for different temperatures $T \approx T_q$. For comparison, the value of the critical bubble solution at $\bar{r} = 0$ and $T = T_q$ is shown by the purple dashed horizontal line.
   }
    \label{fig:near_Tq}
\end{figure}

In this section we analyse in more detail the thermal bounce near the quantum-to-classical transition temperature $T_q$. The difference between the thermal bounce and the critical bubble near $T_q$ is amplified for small radion mass, and therefore we consider $m_\vf^2=0.1$ to explore the region $T/T_q\approx 1$.
We find the two solutions using the method described in appendix~\ref{app:Num} and calculate their Euclidean actions.
The ratio of the thermal bounce action $B_{\rm th}$ to the critical bubble action $B_c$ is shown in the left panel of figure~\ref{fig:near_Tq}, where we also indicate the temperature $T_q'$ 
at which an extra negative mode of the critical bubble appears.
Notice that although the difference is small, $T_q'<T_q$, where $T_q$ is defined as the temperature where $B_{\rm th}=B_c$.
Furthermore, the thermal bounce solution exists for temperatures $T>T_q$ where $B_{\rm th}>B_c$, 
implying that the thermal bounce solution does not branch off from the critical bubble at $T=T_q$. This is explicitly seen in the right panel of figure~\ref{fig:near_Tq}, which depicts the thermal bounce profile at $\br=0$ for several values of $T$ near $T_q$. 
As $T$ increases beyond $T_q$, the time dependence of the bounce diminishes, suggesting that the bounce might still merge with the critical bubble at a higher temperature. 
However, due to the limiting precision of the numerical procedure, we were not able to definitively confirm this behavior.

Nevertheless, these findings indicate that for 
$T>T_q'$ (including temperatures near $T_q$)
the two solutions are different saddle points of the free energy, separated by a barrier in configuration space. For $T<T_q'$ the barrier disappears, and one can flow from the critical bubble to the thermal bounce along the negative mode of the critical bubble. Thus, given that the difference between $T_q'$ and $T_q$ is small even for $m_\vf^2=0.1$,
we will continue to approximate $T_q$ with $T_q'$ and identify $T_q'$ with $T_q$ throughout the text.

\bibliographystyle{JHEP}
\bibliography{Refs}

\end{document}